\documentclass[pra,twocolumn]{revtex4-1}
\usepackage{float,graphicx,amsmath,array}
\begin{document}
\title{ Experimental Realization of Secure Multiparty Quantum Summation Using Five-Qubit IBM Quantum Computer on Cloud}
\author{Ayan Majumder}
\email{ayanmajumder123@gmail.com}
\affiliation{Department of Physical Sciences,
Indian
Institute of Science Education and
Research, Mohali, Punjab-140306, India.}
\author{Santanu Mohapatra}
\email{mohapatrasantanu5@gmail.com}
\affiliation{Department of Physics, Indian Institute of Technology, Kharagpur-721302, India.
}
\author{Anil Kumar}
\email{anilnmr@iisc.ac.in}
\affiliation{
Department of Physics, NMR Research Centre, Indian Institute of Science, Bangalore-560012, India.}
\begin{abstract} A scheme for \textit{secure multiparty quantum summation} was proposed by \textit{Run-hau Shi et al}.,(SCIENTIFIC REPORTS, 6:19655, DOI:10.1038/srep19655). IBM Corporation has
released a superconductivity based 5-qubit quantum computer named \textit{``Quantum Experience'' } and placed it on \textit{``cloud''}. In this paper we take advantage of the online availability of this real quantum processor(ibmqx2) and carry out the above protocol that has never been experimentally demonstrated earlier. Here, we set up experiments for secure quantum summation of one-qubit secret states . Besides, we propose a protocol for square and cubic summation and also simulate the proposed protocol for both cases. Experimental implementation of secure multiparty quantum summation protocol for two or more qubit secret state and the proposed protocol for square(and cubic) summation are not possible using the available 5-qubit quantum processor(ibmqx2). IBM has recently placed a 16-qubit quantum processor(ibmqx3) on the cloud but it is still not accessible for all users. We describe a simulation for two and three qubit secret state and proposed protocol for square(and cubic) summation using a simulator(Custom Topology) which is provided by IBM Corporation.
\end{abstract}
\maketitle
\section{Introduction}
\label{intro}
In our everyday lives, we generate and keep a lot of data which are highly confidential. Secure Multiparty Computation (SMC) plays an important role in Cryptography\cite{Christof Paar and Jan Pelzl}, makes it possible to execute specific programs on confidential data, while ensuring that no sensitive information from the data is leaked\cite{P.Kamm}. IBM Corporation has released the Quantum Experience which allows users to access 5-qubit quantum processor(ibmqx2). We take advantage of the online availability of this real hardware and present the secure multiparty quantum summation. Here, we experimentally implement a protocol which is given by \textit{Run-hau Shi et al.}\cite{Run-hua Shi}, using the five-qubit superconductivity based quantum computer. We also propose a protocol for square and cubic summation and simulate the proposed protocol for both cases.  IBM quantum processor\cite{User guide of IBM quantum experience} which we have used here, is placed at T.J.Watson lab, York Town, USA. Till now several groups have used the 5-qubit quantum computer to demonstrate various experiments\cite{Mitali Sisodia},\cite{M. Hebenstreit},\cite{Simon J. Devitt},\cite{Rui Li},\cite{Christine Corbett Moran}. Recently, we have also used this processor to implement few experiments(to be placed in arXiv soon). These are non-destructive discrimination of arbitrary orthogonal quantum states and no-hiding theorem\cite{Ayan Majumder}, which were earlier implemented using NMR quantum processors\cite{J. R. Samal},\cite{Manu V S},\cite{Jharana Rani Samal}. Recently several groups have discussed hardware for superconductivity based quantum processor\cite{Jhon Clarke},\cite{D. Rosenberg},\cite{A.D. Corcoles}.
Recently Panigrahi et al. have also placed on arXiv experimental verification of a ``Quantum Cheque'' and ``No-Hiding Theorem'', using ibmqx2\cite{B.K. Behera},\cite{P.K. Panigrahi}.

\section{\label{sec:level1}Protocol for secure multiparty quantum summation}
Let, there be m parties; ${P_1},{P_2},......,{P_m}$(where $m>2$) and each party ${P_i}$(where  $1\leq i\leq m$) has a secret 
integer $y_i \in \{0,1,.....,N-1\}$ where, $N=2^n$. All parties jointly compute the summation $S=\displaystyle\sum_{i=1}^{m} {y_i}mod N$
without disclosing their respective secret $y_i$s to each other. According to the reference\cite{Run-hua Shi}, the protocol for the secure multiparty quantum summation is given in Fig.1.\\

\fbox{\includegraphics[width=0.45\textwidth]{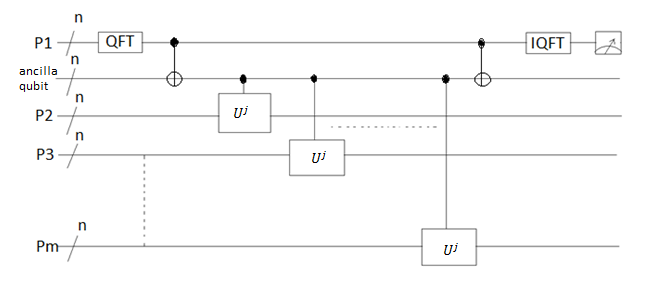}}

\textbf{\textit{Fig.1: Circuit diagram for the protocol for m number of parties and n-qubit case. QFT and IQFT are the Quantum Fourier Transform and Inverse Quantum Fourier Transform respectively. $U^j$'s are the ORACLE operators having n ancilla qubits as the control qubits.}}\\

\begin{itemize}
\item {Let, $P_1$ be the initiator party and  prepares an n-qubit basis state $|y_1\rangle_h$ and this is the private secret of $P_1$. $P_1$ then applies the quantum Fourier transform\cite{Michael A. Nielsen},\cite{Andrew W.} to the state $|y_1\rangle_h$ and gets the following state,(Eq.1)}
\begin{equation} 
 |\phi_1\rangle=QFT|y_1\rangle_h=\frac{1}{\sqrt{2^n}}\displaystyle\sum_{j=0}^{2^n -1} e^{2\pi i \frac{y_1}{2^n}j}|j\rangle_h
\end{equation}

\item {After that, $P_1$ prepares a n-qubit ancillary state ${|0\rangle_t}^{\bigotimes n}$ and performs n-CNOT gate operations on the product state $|\phi_1\rangle {|0\rangle_t}^{\bigotimes n}$,
where each qubit of the first n-qubits is the control qubit and the corresponding qubit of second n-qubit is the target qubit. Then the resultstant state is (Eq.2)}
\begin{equation}
|\phi_2\rangle=CNOT^{\bigotimes n}|\phi_1\rangle {|0\rangle_t}^{\bigotimes n} =\frac{1}{\sqrt{2^n}}\displaystyle\sum_{j=0}^{2^n -1} e^{2\pi i \frac{y_1}{2^n}j}|j\rangle_h |j\rangle_t
\end{equation}

where, the subscript $h$ denotes that the qubit will stay at home and $t$ denotes that the qubit will be transmitted through the authentic quantum channel.
\item Then $P_1$ sends the $|j\rangle_t$ state to $P_2$ through the authentic quantum channel.
\item After getting this $|j\rangle_t$ state, $P_2$ prepares a secret state $|y_2\rangle$ and applies an \textit{oracle}\cite{Run-hua Shi},\cite{Michael A. Nielsen} operator $C_j$(where, $C_j:|j\rangle_t|y_2\rangle \rightarrow |j\rangle_t U^{j}|y_2\rangle$ and $U|y_2\rangle=e^{2\pi i\frac{y_2}{2^n}}|y_2\rangle$) on the 
product state $|j\rangle_t |y_2\rangle$, the total resultant state is (Eq.3),
\[|\phi_3\rangle=C_j \frac{1}{\sqrt{2^n}}\displaystyle\sum_{j=0}^{2^n -1} e^{2\pi i \frac{y_1}{2^n}j}|j\rangle_h |j\rangle_t |y_2\rangle \]
\begin{equation}
=\frac{1}{\sqrt{2^n}}\displaystyle\sum_{j=0}^{2^n -1} e^{2\pi i \frac{y_1 + y_2}{2^n} j}|j\rangle_h |j\rangle_t |y_2\rangle.
\end{equation}

\item After that, $P_2$ sends this $|j\rangle_t$ to $P_3$ through the authentic quantum channel and $P_3$ also follows the same protocol which
$P_2$ followed. This protocol is repeated $m-1$ times. If everyone honestly follows the protocol, the total resultant state will be (Eq.4),
\begin{equation}
|\phi_4\rangle=\frac{1}{\sqrt{2^n}}\displaystyle\sum_{j=0}^{2^n -1} e^{2\pi i [{\sum_{i=1}^{m} y_i}/{2^n}]j}|j\rangle_h |j\rangle_t |y_2\rangle .... |y_m\rangle .
\end{equation}

Then, $P_m$ sends the $|j\rangle_t$ state to $P_1$ through the authentic quantum channel.
\item After getting the $|j\rangle_t$ state, $P_1$ applies the n-CNOT gate on this state to check that there is any dishonest person or not.
If everyone follows the protocol honestly, $P_1$ will get $|0\rangle^{\bigotimes n}$ state. Now the resultant state is (Eq.5),
\begin{equation}
|\phi_5\rangle=\frac{1}{\sqrt{2^n}}\displaystyle\sum_{j=0}^{2^n -1} e^{2\pi i [{\sum_{i=1}^{m} y_i}/{2^n}]j}|j\rangle_h |0\rangle_t |y_2\rangle .... |y_m\rangle .
\end{equation}

\item At last, $P_1$ applies the inverse quantum Fourier transformation\cite{Run-hua Shi},\cite{Michael A. Nielsen},\cite{Andrew W.} to get the result of summation as (Eq.6) 
\begin{equation}
S=\displaystyle\sum_{i=1}^{m} {y_i}mod N.
\end{equation}.

\end{itemize} 
\section{Experimental verification}
\subsubsection{For three party $(m=3)$ and one qubit $(n=1)$ secret state, assume that secret integers for $P_1$, $P_2$ and $P_3$ are 0,1 and 0 respectively.}

\fbox{\includegraphics[width=0.45\textwidth]{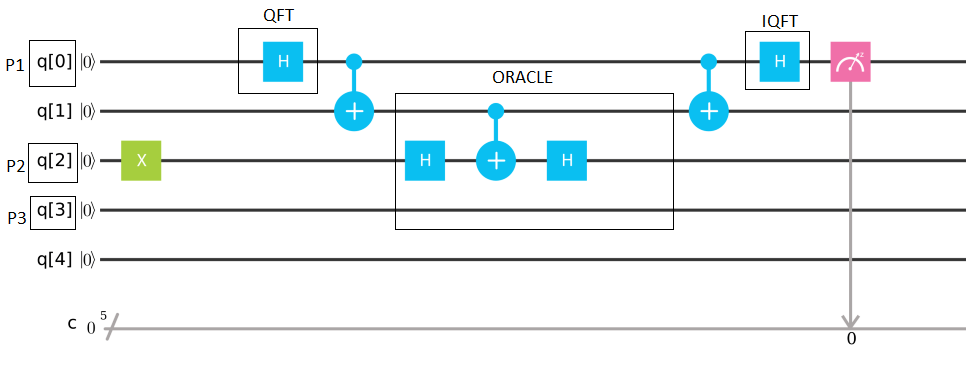}}

\textbf{\textit{Fig.2: Figure showing the experimental implementation of the protocol for m=3 and n=1. $q[1]$ is the ancilla qubit.}}\\

According to the protocol, Hadamard gate is taken to be the quantum Fourier transform\cite{Michael A. Nielsen},\cite{Andrew W.} (Fig. 2). After this, $P_1$ applies the controlled-not gate on the ancilla qubit and sends this qubit($q[1]$) to $P_2$ through the authenticated quantum channel.  Then $P_2$ makes secret state $|1\rangle$ and applies the controlled-U (where $q[1]$ is the control bit and $q[2]$ is the target bit), yielding (Eq.7) 
\begin{equation}
controlled\ U=H_{2}\bigotimes CNOT_{1,2}\bigotimes H_{2}.
\end{equation} 

After that, $P_2$ iterate this operator(means applies controlled-$U^2$ and controlled-$U^4$ respectively, those are $4\times 4$ identity operator). $P_2$ then sends the $q[2]$ qubit to $P_3$ using the authenticated quantum channel. Similarly, $P_3$ makes own secret state $|0\rangle$ and applies the controlled-$U^j$ (where $q[1]$ is the control bit and $q[3]$ is the target bit)(Eq.8). \\

Here, 
\begin{equation}
controlled\ U=controlled\ identity_{1,3}.
\end{equation}

Therefore, overall controlled-U operations can be realized by the optimized \textit{oracle}\cite{Run-hua Shi},\cite{Michael A. Nielsen} operator given in Fig 2.\\

After that, $P_2$ sends the ancillary qubit $q[1]$ to $P_1$ through the authenticated quantum channel. After getting the ancillary state, $P_1$ again applies controlled-not gate on $q[1]$ qubit and measures the $q[1]$ in the computational basis. Here we assume that all parties are honest, therefore $P_1$ will get the $|0\rangle$ state when he measure the q[1]. Then $P_1$ will apply the IQFT\cite{Run-hua Shi},\cite{Michael A. Nielsen},\cite{Andrew W.} operation on his qubit and get the result of summation. Theretically, the summation would be(Eq.9), 
\begin{equation}
S=\displaystyle\sum_{i=1}^{3} {y_i}mod 2 = (0+1+0)mod 2 = 1
\end{equation}

The result of this experiment using the IBM quantum computer is given in Fig. 3,\\

\fbox{\includegraphics[width=0.45\textwidth]{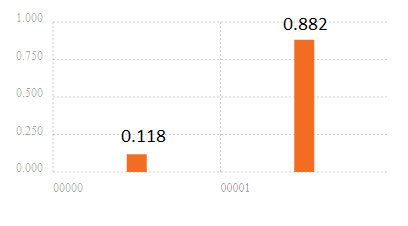}}

\textbf{\textit{Fig.3: Figure showing the experimental result of the protocol for $m=3$ and $n=1$. The above result is obtained by taking an average of 8192 number of shots. The state 1 is obtained with a probability  $0.882 $.}}\\

The probability improves as the number of shots is increased. For example, for 1024 shots, we obtained a probability of 0.863(data not shown).

\subsubsection{For four party $(m=4)$ and one qubit $(n=1)$ secret state, assume that secret integers for $P_1$, $P_2$, $P_3$ and $P_4$ are 1,1,0 and 0 respectively.}

\fbox{\includegraphics[width=0.45\textwidth]{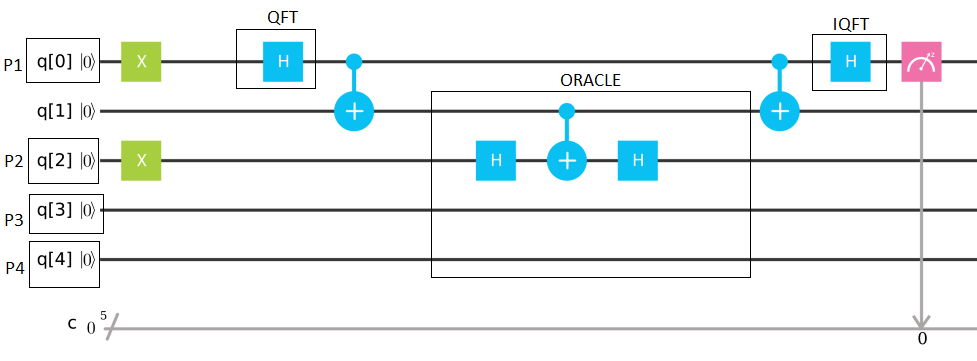}}

\textbf{\textit{Fig.4: Figure showing the experimental implementation of the protocol for $m=4$ and $n=1$. Here, $q[1]$ is the ancilla qubit.}}\\

The \textit{oracle}\cite{Run-hua Shi},\cite{Michael A. Nielsen} operator for ${P_2}$ is also given as eq.[7]. The \textit{oracle}\cite{Run-hua Shi},\cite{Michael A. Nielsen} operator for ${P_3}$ is given as eq.[8], and the \textit{oracle}\cite{Run-hua Shi},\cite{Michael A. Nielsen} operator for ${P_4}$ is given by(Eq.10),
\begin{equation}
controlled\ U=controlled\ identity_{1,4}.
\end{equation}

Thus, overall controlled-U operation can be realized by the optimized \textit{oracle} operator given in Fig 4.\\
 The final result that $P_1$ should get is (Eq.11), \\

\begin{equation}
S=\displaystyle\sum_{i=1}^{4} {y_i}mod 2 = (1+1+0+0)mod 2 = 0.
\end{equation}

The experimental result is given in Fig.5,\\

\fbox{\includegraphics[width=0.45\textwidth]{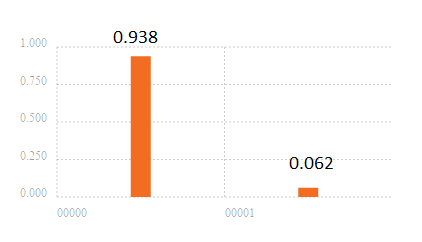}}

\textbf{\textit{Fig.5: Figure showing the experimental result of the protocol for $m=4$ and $n=1$. The above result is obtained by taking an average of 8192 number of shots. The state 0 is obtained with a probability  $0.938 $.}}

\subsubsection{For four party $(m=4)$ and one qubit $(n=1)$ secret state, assume that secret integers for $P_1$, $P_2$, $P_3$ and $P_4$ are 1,1,0 and 1 respectively.}

\fbox{\includegraphics[width=0.45\textwidth]{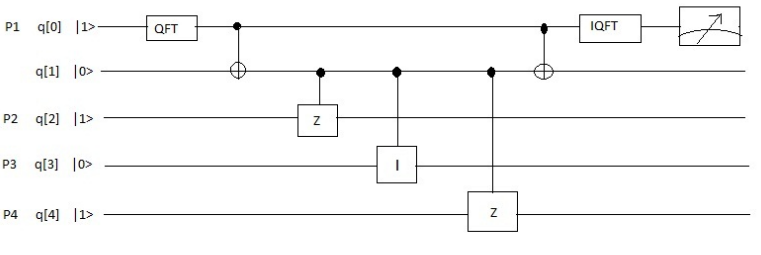}}

\textbf{\textit{Fig.6: Figure showing the experimental implementation of the protocol for $m=4$ and $n=1$. $q[1]$ is the ancilla qubit.}}\\
 
In this experiment, while implementing the optimized oracle operator controlled-Z(given in Fig.7), we needed a controlled-not operation taking q[1] as a control qubit and q[4] as a target qubit. But according to the coupling map of this ibmqx2 processor given in the figure 8. We cannot construct the \textit{oracle}\cite{Run-hua Shi},\cite{Michael A. Nielsen} operator directly. So, we modify the oracle operator in this experiment.

\fbox{\includegraphics[width=0.45\textwidth]{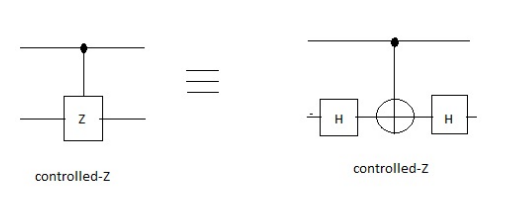}}

\textbf{\textit{Fig.7: Decomposition of controlled-Z operator using the available quantum gates(H and CNOT gate) in ibmqx2 processor.}}\\

\fbox{\includegraphics[width=0.4\textwidth]{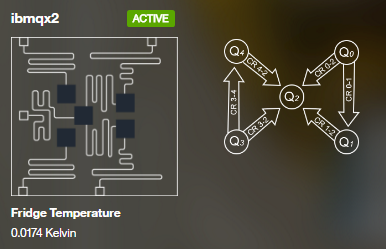}}

\textbf{\textit{Fig.8: ibmqx2 processor\cite{User guide of IBM quantum experience}. Coupling map $= \{0: [1, 2], 1: [2], 3: [2, 4], 4: [2]\}$ where, a: [b] means a CNOT with qubit a as control and b as target can be implemented.}}\\

The \textit{oracle}\cite{Run-hua Shi},\cite{Michael A. Nielsen} operator for ${P_2}$ is given by(Eq.12),
\[controlled\ U\]
\begin{equation}
=H_{1}H_{1}\bigotimes H_{2}\bigotimes CNOT_{1,2}\bigotimes H_{2}\bigotimes H_{1}H_{1} .
\end{equation} 

For $P_3$ the oracle operator is a controlled-identity operator. The \textit{oracle}\cite{Run-hua Shi},\cite{Michael A. Nielsen} operator for ${P_4}$ is given by(Eq.13),
\[controlled\ U\]
\begin{equation}
=H_{4}H_{4}\bigotimes H_{2}\bigotimes CNOT_{4,2}\bigotimes H_{2}\bigotimes H_{4}H_{4} .
\end{equation}

Therefore, overall controlled-U operation can be realized by the optimized \textit{oracle} operator given in Fig.9.\\

\fbox{\includegraphics[width=0.45\textwidth]{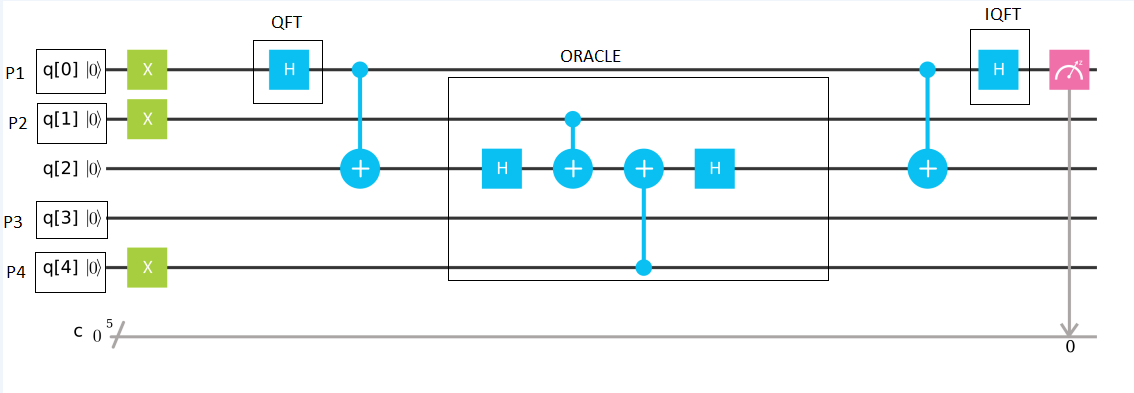}}

\textbf{\textit{Fig.9: Figure showing the modified oracle operator of the experimental implementation of the protocol for $m=4$ and $n=1$. $q[2]$ is the ancilla qubit.}}\\

The measurement result would be (Eq.14),
\begin{equation}
S=\displaystyle\sum_{i=1}^{4} {y_i}mod 2 = (1+1+0+1)mod 2 = 3 mod 2 = 1 .
\end{equation}

The experimental result from the 5-qubit quantum computer is given in Fig.10, \\

\fbox{\includegraphics[width=0.45\textwidth]{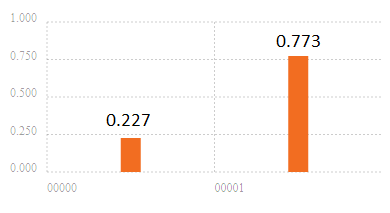}}

\textbf{\textit{Fig.10: Figure showing the experimental result of the protocol for $m=4$ and $n=1$. The above result is obtained by taking an average of 8192 number of shots. The state 1 is obtained with a probability  $0.773$.}}\\

The fidelity\cite{Mitali Sisodia} of the $1^{st}$, $2^{nd}$ and $3^{rd}$ experiment are $93.91\%$, $ 96.85\%$, and $47.64\%$ respectively. For fidelity calculation we use the equation-(15) The lower fidelity of the $3^{rd}$ experiment is due to the incomplete coupling map(Fig.8) leading to more number of operators as given in Fig.9. We also calculate ``average absolute deviation($\langle \Delta x \rangle$)'' and ``maximum absolute deviation($\Delta x_{max}$)'' using equation-(16),(17) for the $1^{st}$, $2^{nd}$ and $3^{rd}$ experiment. The summary of the result of the experiments is given in the table-1 and the state tomography\cite{Mitali Sisodia} of the experimental density matrix of each experiment are given in Fig.11,12,13.
 
\begin{equation}
Fidelity=Tr \sqrt{\sqrt{\rho^T}.\rho^E .\sqrt{\rho^T}}
\end{equation} 
In equation-(15) $\rho^T$ is the theoretical density matrix and $\rho^E$ is the experimental density matrix. 
\begin{equation}
\langle \Delta x \rangle = \frac{1}{N^2}\sum_{i,j=1}^{N} |x^T_{i,j}-x^E_{i,j}|
\end{equation}

\begin{equation}
\Delta x_{max} = Max|x^T_{i,j}-x^E_{i,j}| 
\end{equation}

\[\forall i,j \in \{1,N\}\]

In equations-(16),(17)$x^T_{i,j}$  is the element of theoretical density matrix and $x^E_{i,j}$ is the element of experimental density matrix.
 
\fbox{\includegraphics[width=0.45\textwidth]{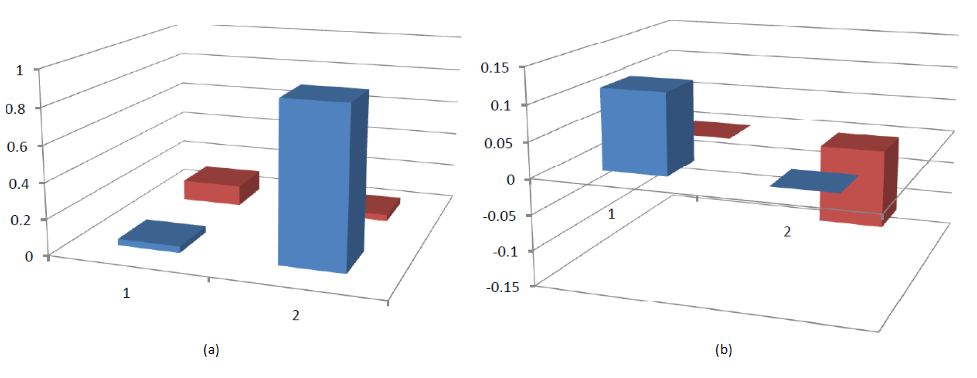}}

\textbf{\textit{Fig.11: The left part and right part of the figure show the real and imaginary part of the experimental density matrix of the experiment-1 respectively.}}

\fbox{\includegraphics[width=0.45\textwidth]{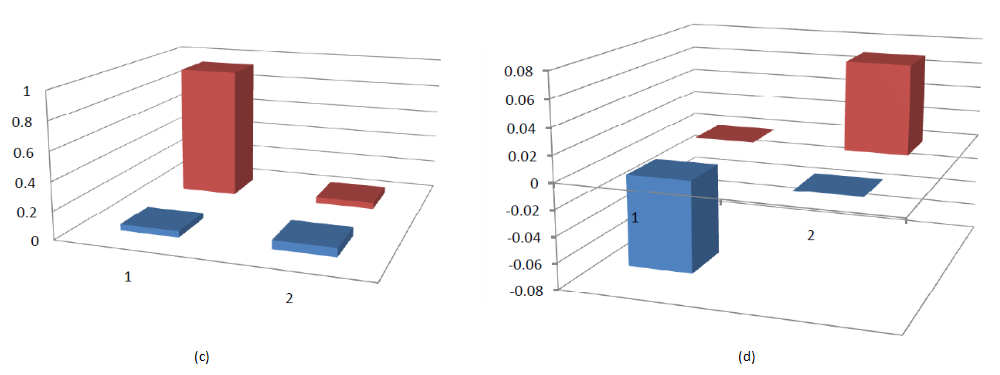}}

\textbf{\textit{Fig.12: The left part and right part of the figure show the real and imaginary part of the experimental density matrix of the experiment-2 respectively.}}

\fbox{\includegraphics[width=0.45\textwidth]{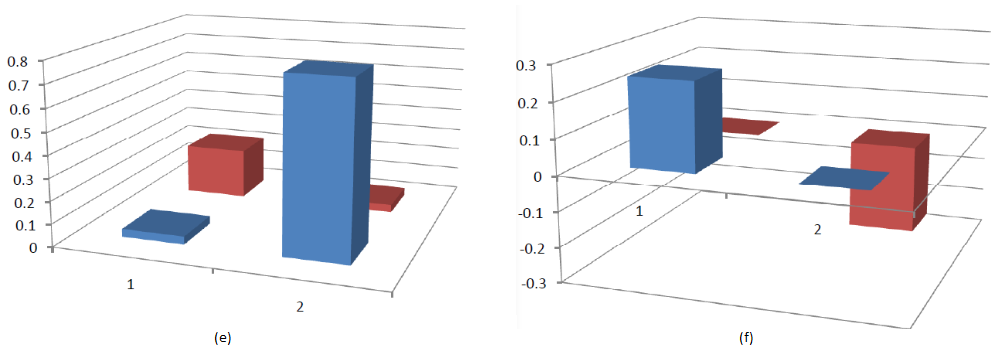}}

\textbf{\textit{Fig.13: The left part and right part of the figure show the real and imaginary part of the experimental density matrix of the experiment-3 respectively.}}

\begin{center}
\begin{tabular}{ | m{2cm} | m{2cm} | m{2cm} | m{2cm} |} 
\hline
Expt. No. & $\langle \Delta x \rangle$ & $\Delta x_{max}$ & Fidelity \\ 
\hline
 1 & 11.92\% & 12.05\% & 93.91\%\\ 
\hline
 2 & 7.23\% & 8.27\% & 96.85\%\\ 
\hline
 3 & 24.12\% & 25.54\% & 47.64\%\\
\hline 
\end{tabular}
\end{center}

\textbf{\textit{Table-1: This table shows the average absolute deviation($\langle \Delta x \rangle$), maximum absolute deviation($\Delta x_{max}$) and the fidelity of the experiments.}}

\section{Simulation of the protocol for the generalized case for $n \geq 2$}

In this case, according to the protocol, number of parties$(m)$ should be greater than two. In the ibmqx2 processor, the number of qubits is five. For one qubit secret state $(y_i \in \{0,1\})$ case we can consider maximum 4-parties, because out of five qubit one should be ancilla qubit. Consider two qubit secret state $(y_i \in \{0,1,2,3\})$, minimum number of parties should be three and two qubit ancilla state, then the total number of qubits should be eight. But, only five qubits are available in the ibmqx2 processor. For this limitation, we are not able to execute this protocol for two or more qubits using this processor. However, there is a quantum simulator available in IBM Quantum Experience namely Custom Topology, which we use to simulate this protocol for two and three qubit secret integer, and more number of parties for one qubit secret integer. 
In the next subsections(A,B and C), we show the simulations. Unlike experiments ``simulation results'' appear with probability 1.

\subsection{For three party $(m=3)$ and two qubit $(n=2)$ secret integer}

\fbox{\includegraphics[width=0.45\textwidth]{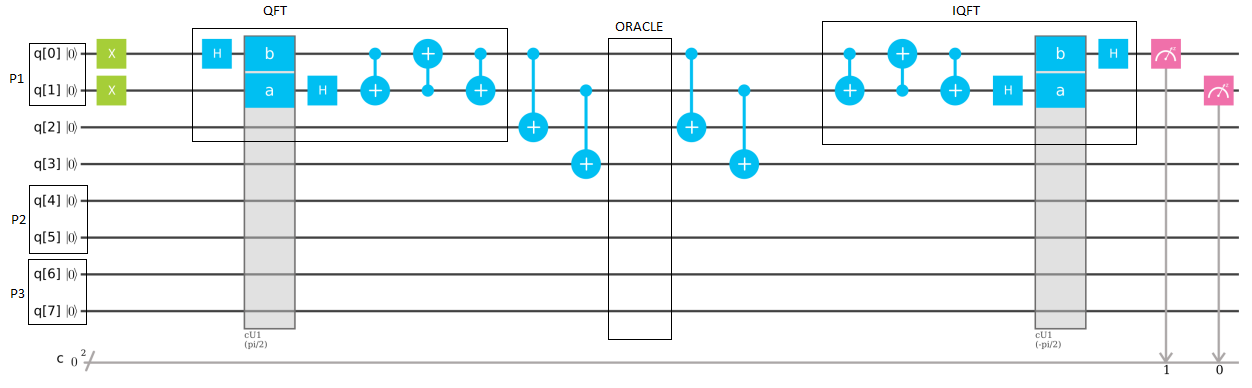}}

\textbf{\textit{Fig.14: Figure showing the simulation of the protocol for $m=3$ and $n=2$. The  $y_i$s are taken to be 3, 0 and 0 respectively and $q[2]$,  $q[3]$ are the ancilla qubits.}}\\

According to the procedure, we apply the QFT and CNOT$^{\bigotimes 2}$ gate respectively. Here, the oracle operators are identity operators.
So, the measurement result should be(Eq.18)

\begin{equation}
S=\displaystyle\sum_{i=1}^{3} {y_i}mod 4 = (3+0+0)mod 4 = 3 
\end{equation}

The simulated result is given in Fig.15,\\

\fbox{\includegraphics[width=0.45\textwidth]{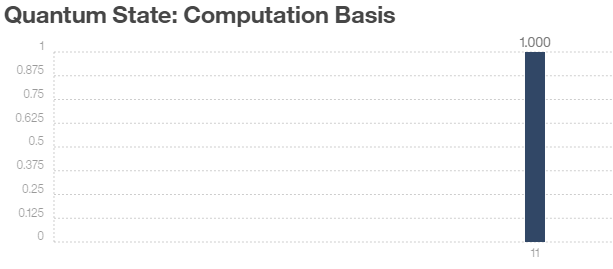}}

\textbf{\textit{Fig.15: Figure showing the simulated result of the protocol for $m=3$ and $n=2$. The state 11(binary representation of 3) is obtained with probability 1.0.}}

\subsection{For three party $(m=3)$ and three qubit $(n=3)$ secret integer}

\fbox{\includegraphics[width=0.45\textwidth]{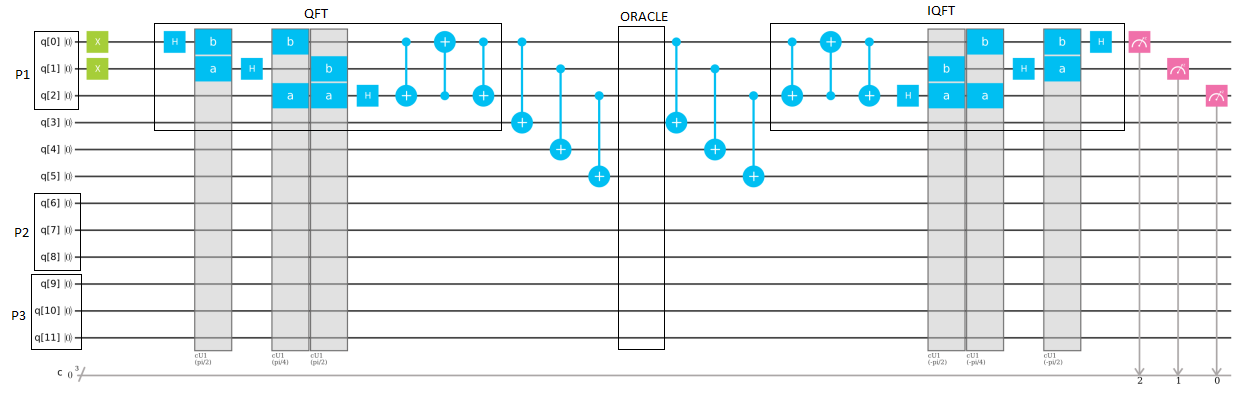}}

\textbf{\textit{Fig.16: Figure showing the simulation of the protocol for $m=3$ and $n=3$. The $y_i$s are taken to be 6, 0 and 0 respectively and $q[3]$, $q[4]$, $q[5]$ are the ancilla qubits.}}\\

Here, oracle operators are identity operators.
So, the measurement result should be(Eq.19)
\begin{equation}
S=\displaystyle\sum_{i=1}^{3} {y_i}mod 8 = (6+0+0)mod 8 = 6
\end{equation}

The simulated result is given in Fig.17,\\

\fbox{\includegraphics[width=0.45\textwidth]{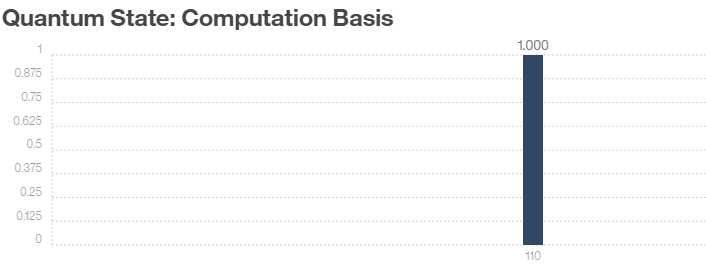}}

\textbf{\textit{Fig.17: Figure showing the simulated result of the protocol for $m=3$ and $n=3$. The state 110(binary representation of six) with probability 1.0.}} 

\subsection{For fifteen party $(m=15)$ and one qubit $(n=1)$ secret integer}

\fbox{\includegraphics[width=0.45\textwidth]{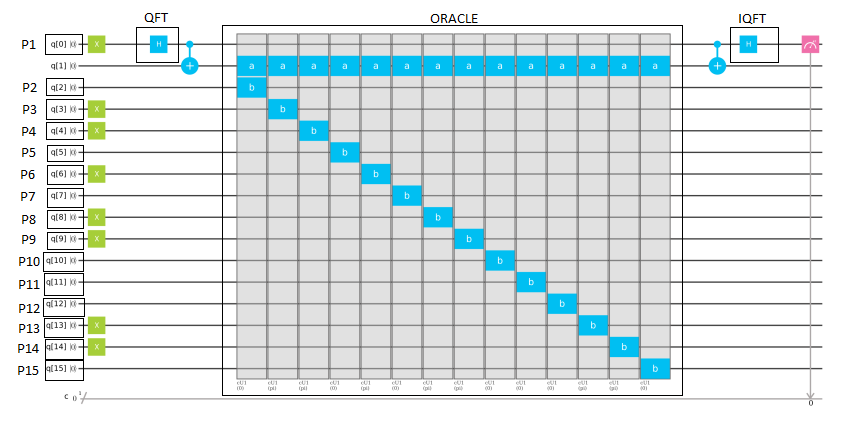}}

\textbf{\textit{Fig.18: Figure showing the simulation of the protocol for $m=15$ and $n=1$. The $y_i$s are taken to be 1, 0, 1, 1, 0, 1, 0, 1, 1, 0, 0, 0, 1, 1 and 0 respectively and $q[1]$ is the ancilla qubit.}}\\

Here, the oracle operator for $P_2, P_5, P_7, P_{10}, P_{11}, P_{12}$ and $P_{15}$ is controlled-identity operator. Oracle operator for others is controlled-Z operator.
So, the measurement result should be(Eq.20),

\[S=\displaystyle\sum_{i=1}^{15} {y_i}mod 2\]
\begin{equation}
=(1+0+1+1+0+1+0+1+1+0+0+0+1+1+0)mod 2 = 0.
\end{equation}

The simulated result is given in Fig.19,\\

\fbox{\includegraphics[width=0.45\textwidth]{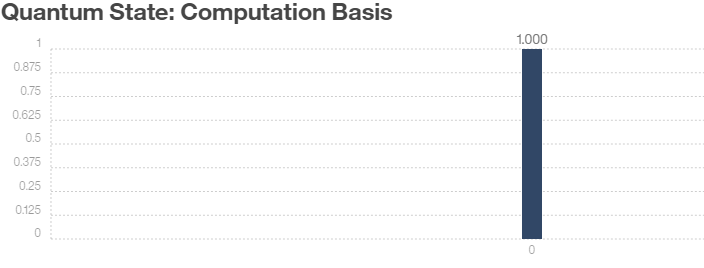}}

\textbf{\textit{Fig.19: Figure shows the simulated result of the protocol for $m=15$ and $n=1$. The state 0 is obtained with probability 1.0.}}

\section{protocol for square and cubic summation}

Here, we propose a protocol for square summation using the protocol given in\cite{Run-hua Shi} and this protocol is as secure as the protocol\cite{Run-hua Shi} for integers summation. Using this proposed protocol, we can compute the summation is given by (Eq.21), 
\begin{equation}
\Gamma=\displaystyle\sum_{i=1}^{m} {y_i^2}mod N,
\end{equation} 

where $y_i \in \{ 0,1,2,3,......N-1 \}$ and $N=2^n$. In this case each party will choose a secret integer $y_i$ and compute ($y_i^2 \ mod \ N$). Then, they make their binary string and execute the above protocol\cite{Run-hua Shi}. We show a simulation for three parties and one qubit secret state. The simulation is given in Fig.20:\\

\fbox{\includegraphics[width=0.45\textwidth]{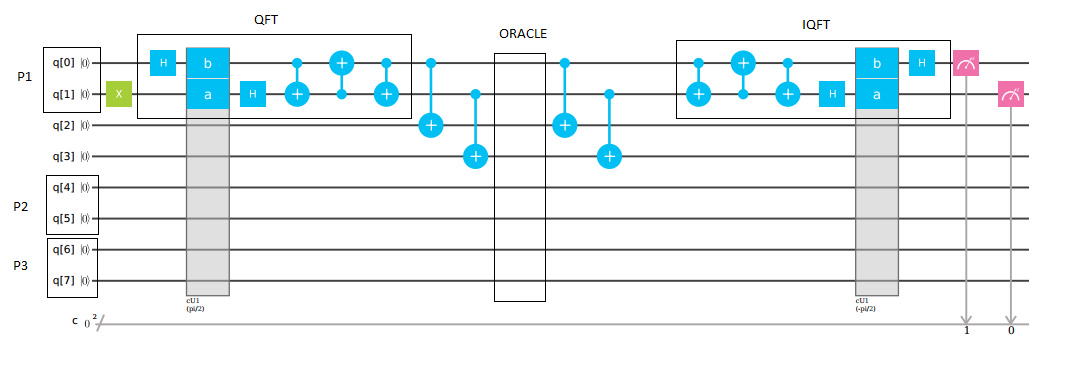}}

\textbf{\textit{Fig.20: Circuit diagram for the square summation protocol.}}\\

Here, $P_1$,$P_2$ and $P_3$ choose the secret integers 3,2 and 0 respectively. The oracle operators are identity operator. So, the square summation(Eq.22),
\begin{equation}
\Gamma=\displaystyle\sum_{i=1}^{m} {y_i^2}mod N = \displaystyle\sum_{i=1}^{3} {y_i^2}mod 4 = (9+4+0)mod4 = 1.
\end{equation}

According to the protocol, $P_1$,$P_2$ and $P_3$ compute ($y_i^2 \ mod \ N$) and make their binary string and execute the \textit{secure multiparty quantum summation}. The result of this simulation is given in Fig.21,\\

\fbox{\includegraphics[width=0.45\textwidth]{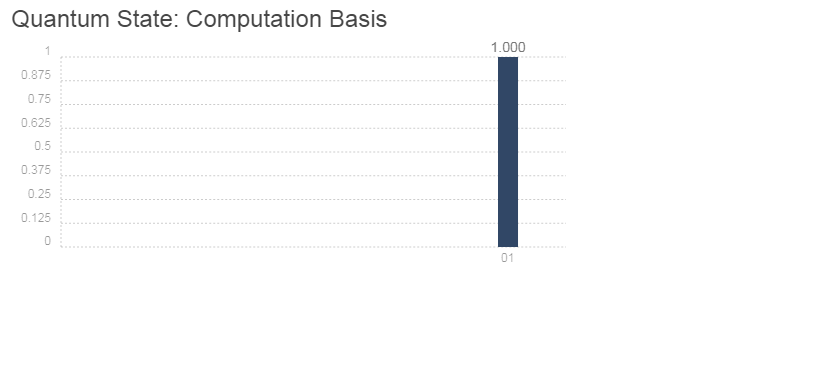}}

\textbf{\textit{Fig.21: Figure showing the simulated result of the proposed protocol for square summation. The state 01(binary representation of 1) is obtained with probability 1.0.}}\\

This proposed protocol for square summation isn't bounded within the square only. Using the same protocol we can compute summation for cubes and higher powers. 
The simulation for cubic summation is given in Figs.22 and 23,\\

\fbox{\includegraphics[width=0.45\textwidth]{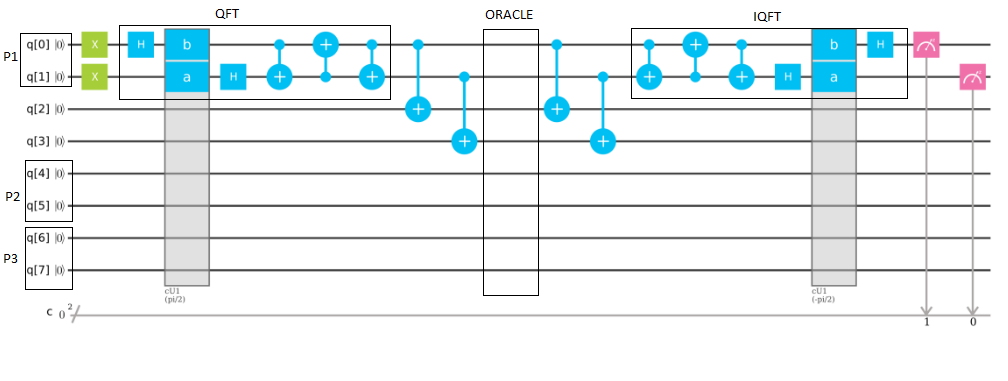}}

\textbf{\textit{Fig.22: Circuit diagram for the cubic summation protocol.}}\\

Here, $P_1$,$P_2$ and $P_3$ choose the secret integers 3,2 and 0 respectively. According to the protocol, $P_1$,$P_2$ and $P_3$ compute ($y_i^3 \ mod \ N$) and make their binary string and execute the \textit{secure multiparty quantum summation}. The result should be (Eq.23),
\begin{equation}
\Omega=\displaystyle\sum_{i=1}^{m} {y_i^3}mod N = \displaystyle\sum_{i=1}^{3} {y_i^3}mod 4 = (27+8+0)mod4 = 3.
\end{equation}

Here, the oracle operators for $P_2$ and $P_3$ are identity operator.\\

\fbox{\includegraphics[width=0.45\textwidth]{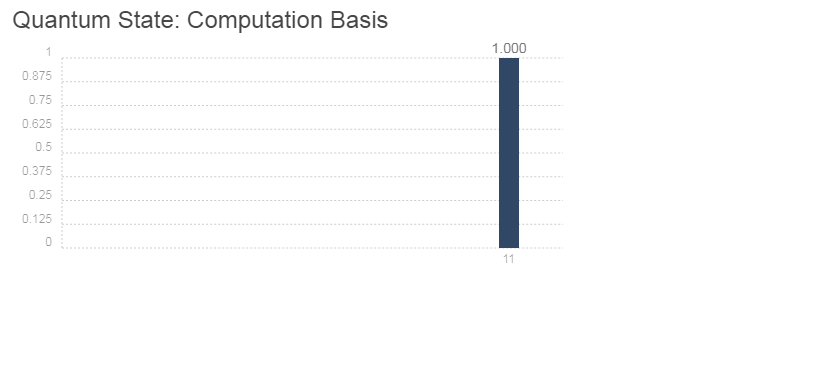}}

\textbf{\textit{Fig.23: Figure showing the simulated result of the proposed protocol for cubic summation. The state 11(binary representation of 3) is obtained with probability 1.0.}}\\

\section{Conclusion}

A protocol for secure multiparty quantum summation has been described\cite{Run-hua Shi} and experimentally implemented here for one qubit secret integer using the five-qubit superconductivity\cite{Jhon Clarke},\cite{D. Rosenberg},\cite{A.D. Corcoles} based quantum computer of IBM Corporation placed on the cloud named ``Quantum Experience''. \\
The generalization of this protocol for summation of 2 $\&$ 3 qubit secret state and square and cubic summation are demonstrated  using Custom Topology (a quantum simulator of IBM Quantum Experience). Due to the limitations of the ibmqx2 processor, it is not possible to experimentally execute this protocol for two and three qubit secret state. However, IBM Corporation has already made a sixteen-qubit processor(ibmqx3). Using this ibmqx3 processor, one can overcome those limitations.  However, till now, it is not available for all users. 

\section{Acknowledgement}
 
Thanks are due to \textit{Prof. Apoorva Patel} of \textit{Indian Institute of Science, Bangalore} and \textit{Dr. K.V.R.M. Murali} member of \textit{ Vijna Labs} for advise and encouragement. 

%
\end{document}